\documentclass[a4paper,12pt]{article}
\pdfoutput=1
\usepackage{graphicx, rotating,amssymb}
\ifx\pdfoutput\undefined
\usepackage[dvips,bookmarks]{hyperref}	% This is for arXiv.org
\else
\usepackage{hyperref}	% This is for pdftex
\fi
\hypersetup{colorlinks,bookmarksopen,bookmarksnumbered,citecolor=verdes,
linkcolor=blus,pdfstartview=FitH,urlcolor=rossos}
\def\myurl#1#2{\href{http://#1}{#2}}
\def\hhref#1{\href{http://arxiv.org/abs/#1}{#1}} % in bibliography
\usepackage{multicol}
\usepackage{amsmath}
\usepackage{color}
\definecolor{rosso}{cmyk}{0,1,1,0.4}
\definecolor{rossos}{cmyk}{0,1,1,0.55}
\definecolor{rossoc}{cmyk}{0,1,1,0.2}
\definecolor{blu}{cmyk}{1,1,0,0.3}
\definecolor{blus}{cmyk}{1,1,0,0.6}
\definecolor{bluc}{cmyk}{1,1,0,0.1}
\definecolor{verde}{cmyk}{0.92,0,0.59,0.25}
\definecolor{verdec}{cmyk}{0.92,0,0.59,0.15}
\definecolor{verdes}{cmyk}{0.92,0,0.59,0.4}

\font\tenrsfs=rsfs10 at 12pt
\font\sevenrsfs=rsfs7
\font\fiversfs=rsfs5
\newfam\rsfsfam
\textfont\rsfsfam=\tenrsfs
\scriptfont\rsfsfam=\sevenrsfs
\scriptscriptfont\rsfsfam=\fiversfs
\def\mathscr#1{{\fam\rsfsfam\relax#1}}

\def\circa#1{\,\raise.3ex\hbox{$#1$\kern-.75em\lower1ex\hbox{$\sim$}}\,}

\newcommand{\beq}{\begin{equation}}
\newcommand{\eeq}{\end{equation}}

\def\circa#1{\,\raise.3ex\hbox{$#1$\kern-.75em\lower1ex\hbox{$\sim$}}\,}
\makeatletter

\def\art{\@ifnextchar[{\eart}{\oart}}
\def\eart[#1]#2#3#4#5#6{{\rm #2}, {#3 #4} {\rm (#6) #5} [{\hhref{#1}}]}
\def\hepart[#1]#2{{\rm #2, \hhref{#1}}}
\newcommand{\oart}[5]{{\rm #1}, {#2 #3} {\rm (#5) #4}}

\newcounter{alphaequation}[equation]
\def\thealphaequation{\theequation\hbox to
0.6em{\hfil\alph{alphaequation}\hfil}}
\def\eqnsystem#1{
\def\@eqnnum{{\rm (\thealphaequation)}}
\def\@@eqncr{\let\@tempa\relax \ifcase\@eqcnt \def\@tempa{& & &} \or
  \def\@tempa{& &}\or \def\@tempa{&}\fi\@tempa
  \if@eqnsw\@eqnnum\refstepcounter{alphaequation}\fi
\global\@eqnswtrue\global\@eqcnt=0\cr}
\refstepcounter{equation} \let\@currentlabel\theequation \def\@tempb{#1}
\ifx\@tempb\empty\else\label{#1}\fi
\refstepcounter{alphaequation}
\let\@currentlabel\thealphaequation
\global\@eqnswtrue\global\@eqcnt=0 \tabskip\@centering\let\\=\@eqncr
$$\halign to \displaywidth\bgroup \@eqnsel\hskip\@centering
$\displaystyle\tabskip\z@{##}$&\global\@eqcnt\@ne
\hskip2\arraycolsep\hfil${##}$\hfil& \global\@eqcnt\tw@\hskip2\arraycolsep
$\displaystyle\tabskip\z@{##}$\hfil
\tabskip\@centering&\llap{##}\tabskip\z@\cr}
\def\endeqnsystem{\@@eqncr\egroup$$\global\@ignoretrue} \makeatother

\begin{document}
\begin{center}
%{XXX-xx/2008-xx}
{ \hfill SACLAY--T09/065}
\color{black}
\vspace{0.5cm}

{\LARGE \bf Constraints on Dark Matter annihilations\\ from reionization and\\[3mm] heating of the intergalactic gas}

\medskip
\bigskip\color{black}\vspace{0.6cm}

{
{\large\bf Marco Cirelli}$^a$,
{\large\bf Fabio Iocco}$^{a,b}$,
{\large\bf Paolo Panci}$^{a,c,d}$
}
\\[7mm]
{\it $^a$ Institut de Physique Th\'eorique, CNRS URA 2306 \& CEA/Saclay,\\ 
	F-91191 Gif-sur-Yvette, France}\\[3mm]
{\it $^b$ Institut d'Astrophysique de Paris, UMR 7095-CNRS and\\ Universit\'e Pierre et Marie Curie, boulevard Arago 98bis, 75014, Paris, France.}\\[3mm]
{\it $^c$ Dipartimento di Fisica, Universit\`a degli Studi dell'Aquila,\\ 
67010 Coppito (AQ), Italy}\\[3mm]
{\it $^d$ Universit\'e Paris 7-Diderot, UFR de Physique,\\ B\^atiment Condorcet, 10 rue A.Domon et L.Duquet, 75205 Paris, France}	
\end{center}

\bigskip

\centerline{\large\bf Abstract}
\begin{quote}
\color{black}
Dark Matter annihilations after recombination and during the epoch of structure formation deposit energy in the primordial intergalactic medium, producing reionization and heating. We investigate the constraints that are imposed by the observed optical depth of the Universe and the measured temperature of the intergalactic gas. We find that the bounds are significant, and have the power to rule out large portions of the `DM mass/cross section' parameter space. The optical depth bound is generally stronger and does not depend significantly on the history of structure formation. The temperature bound can be competitive in some cases for small masses or the hadronic annihilation channels (and is affected somewhat by the details of structure formation). We find in particular that DM particles with a large annihilation cross section into leptons and a few TeV mass, such as those needed to explain the PAMELA and FERMI+HESS cosmic ray excesses in terms of Dark Matter, are ruled out as they produce too many free electrons. We also find that low mass particles ($\lesssim 10$ GeV) tend to heat too much the gas and are therefore disfavored.
\end{quote}

%\tableofcontents
%\newpage

\section{Introduction}
\label{introduction}

After having settled in a state of electrical neutrality at the time of recombination, the atoms of the intergalactic baryonic gas had their electrons stripped off and returned to the state of an ionized optically thick plasma with free electrons during the process which is known as `reionization'~\cite{reviewsreion}. The phenomenological information of how and when this happened comes mainly from two sets of observations, both involving the study of the absorption of photons that have crossed such intergalactic medium (IGM), but of very different energies and in different contexts: (i) the measurement of the optical depth $\tau$ encountered by photons of the Cosmic Microwave Background (CMB) during their journey from the surface of recombination to us and (ii) the analysis of the spectra of distant quasars.\\
In this second class, the Gunn-Peterson test (the absence of complete absorption at Lyman-$\alpha$ frequency) in the spectra of quasars located at redshift $z \lesssim 6$ allows to infer that the process of reionization of all the hydrogen atoms was complete by then~\cite{GP}. Similarly, ultraviolet spectral observations suggest that helium has also been doubly ionized since redshift $z \approx 3$~\cite{HeII}. The resulting free electron population between us and redshift $z \simeq 6$ translates quantitatively into a value for the optical depth $\tau$ of about 0.038 (see Sec.\ref{computing}).\\
The WMAP 5-yr measurements give, however, a higher value~\cite{WMAP5yr}
\beq
\tau = 0.084 \pm 0.016
\label{tauWMAP}
\eeq
(this number is determined combining with Supernovae and Baryon Acoustic Oscillations observations; the CMB data alone give a very similar result: $\tau = 0.087 \pm 0.017$~\cite{WMAP5yrparameters}).
This implies that it must have existed a population of free electrons, at $z \gtrsim 6$, that contribute to the optical depth the remaining $\delta\tau \simeq 0.046  \pm 0.016 \le 0.062$ (at $1\sigma$).

As of now, however, the possible contributions of different kinds of high redshift astrophysical sources to reionization are still a matter of debate: whether first or second generation stars (Population III and II, respectively) or older quasars have given the dominant contribution, and the nature of its evolution with redshift is still unknown, see e.g.~\cite{reiosources} and references therein. 

\medskip

%The same sources that induced the ionization of the IGM must have contributed to its heating as well
Astrophysical sources also heat up the IGM later in the evolution of the Universe: existing measurements of the temperature of the intergalactic medium, therefore, might also allow to place constraints on the kind of sources and their history, see \cite{TigmObs,TigmObs2} and references therein. 
Such temperature is measured via observations of the Ly-$\alpha$ forest, caused by the small amounts of relic neutral gas on the emission lines of distant Quasi Stellar Objects. 
Observations of the Ly-$\alpha$ forest are limited to redshifts $2 \lesssim z \lesssim 5$\,: at higher redshifts the increasing neutral hydrogen density saturates the forest into a complete absorption; at lower redshifts the measurements become too plagued by observational uncertainties.
The determination of the equation of state and temperature of the IGM through Ly-$\alpha$ observations involves also calibrations by means of numerical simulations, and virtually a dependence on cosmology which is taken into account by adding systematics to the observational uncertainties.
We take as a reliable estimation the datapoints in Schaye et al. (2000)~\cite{TigmObs2} (left panel of their Fig. 6). 
These points span $z \simeq 4.3 \to 2$ (see fig.~\ref{results}b) and we will take all of them into account. To fix the ideas, we report here the explicit value of the highest redshift point, which will turn out to be most constraining (and is probably least affected by systematic uncertainties). It reads
\beq
5.1 \cdot 10^3\, {\rm K} \leq T_{\rm igm} \leq 2.0 \cdot 10^4\, {\rm K}
\label{TIGM}
\eeq
at $z = 4.3$ at $1\sigma$ confidence level.

\subsection{The role of Dark Matter}

Dark Matter (DM), which constitutes about 80\% of the total matter in the Universe, can naturally act as a source of ionizing radiation and gas heating. Indeed, most DM candidates are predicted to self-annihilate into Standard Model particles, thus injecting energy into the surrounding medium. From redshift of a few hundred down to now, DM has been gravitationally collapsing into the protohalos and halos that end up forming the cosmic structures observed today; this condensation into bound systems enhances the DM self--annihilation rate (with respect to the contribution of the DM smooth density field) %\footnote{We assume that DM halos do not retain the products of the annihilation process inside them. It can be shown that high energy photons stream easily out of the gas cocoon of the halos where annihilations take place copiously. High energy electrons Inverse Compton upscatter the CMB photons, as we discuss below, and the resulting high energy photons also stream out into the IGM.} 
and therefore can increases significantly the rate of energy deposition, considering that the DM halos do not retain the products of the annihilation process inside the gas cocoon.\\
The ionizations and heating can be produced both by the highly energetic photons directly emitted in the annihilation of two DM particles (with an energy of the order of the DM mass: typically from tens of GeV to tens of TeV) and by the lower energy photons produced by Inverse Compton (IC) scattering of the CMB photons on the energetic $e^+$ and $e^-$ from DM annihilations. The latter turns actually out to be by far the most important process~\cite{BH}: in fact, the cross section for $\gamma e^-$ scattering decreases rapidly with the energy of the impinging photon, so that low energy photons are more efficient in knocking off the electrons from the atoms.\footnote{For instance: for a Dark Matter particle mass of 100 GeV, the prompt photons will have a similar energy while the ones produced by IC scattering from CMB photons at $z \sim 10$ will peak at about 100 MeV; the former ones will have a scattering cross section on the electrons in the atoms of hydrogen of the order of $10^{-29}\ {\rm cm}^2$ while the latter ones of  $10^{-26}\ {\rm cm}^2$, a difference in ionization efficiency of three orders of magnitude. We will discuss the relevant cross sections in detail later.} These `primary reionization' electrons then deposit their energy in the intergalactic medium through several other interactions, freeing many more electrons (that contribute to the optical depth) and also augmenting the temperature of the gas.
 
\medskip

The effect of DM annihilations on reionization has been previously considered in a number of works~\cite{previous, NS, NS2}, that mainly addressed the case of light Dark Matter candidates. More recently, ref.~\cite{BH} has pointed out the importance for reionization of the Inverse Compton photons discussed above, and has studied the possibility of having DM produce the required `missing' reionization $\delta \tau$ discussed above. 

We take the somewhat opposite approach of investigating whether the requirement of not exceeding the optical depth measured by WMAP can impose significant bounds on the properties of DM.
We also consider whether constraints can come from the measurements of the temperature of the intergalactic medium. 

\bigskip

The crucial parameters that determine the amount of (ionizing) energy emitted by DM annihilations are the mass of the DM particle $m_\chi$ (in first approximation, most of its rest mass becomes photons and heat) and the average annihilation cross section $\langle \sigma v \rangle$. Up to one year ago, the benchmark values for these quantities were typically taken to be $m_\chi \simeq$ tens to hundreds of GeV and $\langle \sigma v \rangle_{\rm thermal} \simeq 3\cdot 10^{-26}\ {\rm cm}^3/{\rm sec}$, the values for which the relic abundance of DM particles comes out, via the thermal freeze-out process, to match the observed $\Omega_{\rm DM} h^2 =0.110 \pm 0.005$~\cite{cosmoDM}\footnote{Here $\Omega_{\rm DM} = \rho_{\rm DM}/\rho_c$ is defined as usual as the energy density in Dark Matter with respect to the critical energy density of the Universe $\rho_c = 3 H_0^2/8\pi G_N$, where $H_0$ is the present Hubble parameter. $h$ is its reduced value $h = H_0 / 100\ {\rm km}\, {\rm s}^{-1} {\rm Mpc}^{-1}$.}.\\
Recently, however, larger DM masses (around a few TeV up to multi-TeV) and much larger cross sections (of the order of $10^{-24}\, {\rm cm}^3/{\rm sec}$ up to $10^{-20} \,{\rm cm}^3/{\rm sec}$ or more, depending on the mass of the candidate and the annihilation channel) have been invoked in the hope of explaining in terms of galactic DM annihilations\,\footnote{The origin of these excesses could simply lie in ordinary (albeit possibly peculiar) astrophysical sources, such as one or more pulsars~\cite{pulsars}, sources of CR in galactic spiral arms~\cite{Piran} or secondary production and acceleration of CR in aged SuperNova remnants~\cite{Blasi}. In this case, the sources would not be cosmological but instead located in the galactic disk, and moreover not too far from the Earth, since $e^\pm$ quickly loose energy when travelling from farther away that about 1 kpc. These explanations will be confirmed or ruled out by further, more precise measurements of the spectra and possibly improved computations of the expected yields.} the anomalous signals reported in the fluxes of charged cosmic rays (CR) by several experiments: PAMELA~\cite{PAMELA}, ATIC~\cite{ATIC-2}, PPB-BETS~\cite{PPB-BETS}, FERMI~\cite{FERMIleptons} and HESS~\cite{HESSleptons, HESSleptons2}, as we will discuss in detail below.

The DM interpretation of these signals is already constrained to various degrees by other astrophysical and cosmological bounds: apart from the anti-proton bounds already taken into account~\cite{CKRS, Donatopbar}, limits come from gamma ray and radio waves from the galactic center~\cite{BCST}, from ICS gamma rays from the galactic halo~\cite{CirelliPanci,StrumiaPapucci}, from the integrated cosmological flux of ICS photons~\cite{Profumo,HooperICS}, from neutrinos from the galactic center~\cite{neutrinos}, from CMB observations~\cite{Galli,Slatyer} and from Big Bang Nucleosynthesis~\cite{BBN}. In particular, the works in~\cite{Galli} and~\cite{Slatyer} are closely connected with our analysis, as they also deal with the deposition of energy from DM annihilation and the impact on CMB observables. We will comment later on the complementarity of the studies.

\medskip

Our aim is to confront these recent DM scenarios and, more generally, the framework of annihilating Dark Matter (including the more traditional thermal annihilation cross section and tens-of-GeV mass) with the constraints that can be imposed by its role during the reionization epoch. We find that the bounds are indeed generally very relevant, and have the power to rule out large portions of the $m_\chi-\langle \sigma v \rangle$ parameter space.

\medskip

The rest of the paper is organized as follows. In Sec.\ref{computing} we review the basic formalism which we use to compute the effects of DM annihilation during reionization. In Sec.\ref{discussion} we discuss the implementation of that formalism, the choices of parameters that we adopt and the dependence of the results on such choices. We also present the DM scenarios that we consider and the fits that we perform to the CR data mentioned above. In Sec.\ref{results} we present our results and we discuss the bounds that they produce.

%%%%%%%%%%%%%%%%%%%%%%%
%%%%%%%%%%%%%%%%%%%%%%%

\section{Computing the reionization and heating induced by DM annihilations}
\label{computing}

In this Section we briefly discuss the formalism that we use to compute semi-analytically the effect of DM annihilations during the epoch of reionization.
We follow in particular the discussions in Ref.~\cite{NS,NS2,BH}.

The first quantity that we need to compute is the total optical depth encountered by the CMB photons as they travel from the surface of last scattering to us. It is given, in full generality, by the integral over the time of travel of the photon of the number density of unbound electrons $n_e(z)$ (the scattering targets) multiplied by the Thomson scattering cross section $\sigma_{\rm T} = 8 \pi r_e^2/3 = 0.6652$ barn (in terms of the classical electron radius $r_e$)\footnote{We always work in natural units $c = \hbar = k_B = 1$.}
\beq
\tau = - \int n_e(z)\, \sigma_{\rm T}\, \frac{dt}{dz}.
\eeq
Here 
\beq
\frac{dt}{dz} = - \frac{1}{H_0\, (1+z) \sqrt{\Omega_{\rm M} (1+z)^3 + \Omega_\Lambda}} \simeq - \frac{1}{H_0\,  \sqrt{\Omega_{\rm M}}\, (1+z)^{5/2}}
\label{dtdz}
\eeq
is the standard $\Lambda$CDM relation between time and redshift, approximated for the regime of matter domination in which we are interested. We use the central values inferred from the WMAP-5yr measurements~\cite{WMAP5yr} for the cosmological parameters today: $H_0 = 70.1\, {\rm km}\, {\rm sec}^{-1} {\rm Mpc}^{-1}$, $\Omega_{\rm M} = 0.279$, $\Omega_{\rm b} = 0.0462$ and $\Omega_\Lambda = 0.721$.
As discussed in the Introduction, the totality of the hydrogen and helium gas is assumed to be ionized below redshift 6, and helium is also doubly ionized below redshift $z = 3$. Recalling that helium constitutes about 24\% in mass~\cite{PDG} of the baryonic content of the universe (so that the number of helium atoms $n_{\rm He} = 0.06\, n_{\rm b}$, while for hydrogen $n_{\rm H} = 0.76\, n_{\rm b}$), one can simply express $\tau$ in terms of the number density of atoms today $n_{\rm A} = (0.76+0.06)\, n_{\rm b} = 0.82 \, \rho_c \Omega_{\rm b}/m_p \simeq 1.92\cdot 10^{-7} {\rm cm}^{-3}$ as 
\begin{equation}
\tau =  \underbrace{n_{\rm A} \, \sigma_{\rm T}\, \left[- \frac{0.88}{0.82} \int_0^3 dz  \frac{dt}{dz} (1+z)^3 - \int_3^6 dz  \frac{dt}{dz} (1+z)^3 \right]}_{\mbox{0.038}} + \underbrace{ n_{\rm A} \, \sigma_{\rm T} \left[- \int_6^\infty dz  \frac{dt}{dz} (1+z)^3 x_{\rm ion}(z) \right]}_{\mbox{ $\delta \tau$}}
\label{tau}
\end{equation}
In the above relations, $n_{\rm b}$ and $\Omega_{\rm b}$ represent the number density and energy fraction of baryons today ($m_p$ being the proton mass) and the factors of $(1+z)^3$ rescale the densities to any redshift. 

$\delta \tau$ denotes the amount of early optical depth caused by the unknown fraction $x_{\rm ion}(z)$ of (singly) ionized atoms above redshift 6.
Such reionized fraction obeys the differential equation
\beq
n_{\rm A} (1+z)^3 \frac{dx_{\rm ion}(z)}{dt} = I(z)-R(z),
\eeq
or, equivalently, in terms  of redshift 
\beq
- n_{\rm A} H_0 \sqrt{\Omega_{\rm M}} (1+z)^{11/2} \frac{dx_{\rm ion}(z)}{dz} = I(z)-R(z).
\label{dxdz}
\eeq
On the right hand side are the rate of ionization per volume $I(z)$, that tends to increase $x_{\rm ion}$, and the rate per volume $R(z) = R_{\rm H}(z) + R_{\rm He}(z)$ with which hydrogen and helium atoms of the IGM tend to recombine even while reionization is proceeding. These recombination rates are explicitly given by the following expressions. For hydrogen
\beq
R_{\rm H}(z) = \kappa_{\rm H}\, n_{\rm H}\, n_{e^-} = \kappa_{\rm H} \frac{0.76}{0.82} \left( n_{\rm A} (1+z)^3 x_{\rm ion}(z)\right)^2 
\eeq
where $\kappa_{\rm H} \simeq 3.75 \cdot 10^{-13}  \big(T_{\rm igm}(z)/{\rm eV}\big)^{0.724} {\rm cm^3}/{\rm sec}$ is an effective coefficient determined by fits to experimental data~\cite{Abel}. $T_{\rm igm}(z)$ is the temperature of the IGM, also affected by DM annihilations, that we will discuss below.
Similarly, for helium
\beq
R_{\rm He}(z) = \kappa_{\rm He} \frac{0.06}{0.82} \left( n_{\rm A} (1+z)^3 x_{\rm ion}(z) \right)^2 
\eeq
with $\kappa_{\rm He} \simeq 3.925 \cdot 10^{-13}  \big(T_{\rm igm}(z)/{\rm eV}\big)^{0.635} {\rm cm^3}/{\rm sec}$~\cite{Abel}.

The rate of ionizations per volume produced by DM annihilations at any given redshift $z$ is given by
\beq
I(z) = \int_{e_{\rm i}}^{m_\chi} dE_\gamma \, \frac{dn}{dE_\gamma}(z) \cdot P(E_\gamma, z) \cdot N_{\rm ion}(E_\gamma)
\eeq
where $\frac{dn}{dE_\gamma}(z)$ is the spectral number density of DM-produced photons that are present at redshift $z$, which we will discuss extensively below, and one has to integrate over all photon energies $E_\gamma$ from the H ionization  energy $e_{\rm i}$ (or the He one, we here for simplicity do not distinguish the two) up to the maximum energy $m_\chi$.
$P(E_\gamma, z)$ is the probability of primary ionizations per second, given by 
\beq
P(E_\gamma, z) = n_{\rm A} (1+z)^3 \left[ 1-x_{\rm ion}(z) \right] \cdot \sigma_{\rm tot}(E_\gamma),
\eeq
since the first terms represent the number of target atoms that can be ionized and $\sigma_{\rm tot}$ is total cross section for all  the interactions suffered by the DM-sourced photon and that result in the production of free electrons. It contains several contributions (we follow e.g. the discussion in~\cite{Slatyer}): the cross section for atomic photo-ionization $\gamma A \to e^- A^+$~\cite{ZS} (dominant up to about 1 MeV), the Klein-Nishina cross section for Compton scattering $\gamma e^- \to \gamma e^-$  ~\cite{KN} (dominant to about 1 GeV) and the cross section for pair production on matter $\gamma A \to e^\pm A^\prime$ ~\cite{ZS} (important at energies larger than 1 GeV). 
At higher energies, another processes that produces free electrons becomes important: pair production on CMB photons $\gamma\, \gamma_{\rm CMB} \to e^+ e^-$. At redshift $z \lesssim$ few hundred in which we are interested, its threshold is however above 10 TeV. We do not include the scatterings $\gamma\, \gamma_{\rm CMB} \to \gamma\, \gamma$, as they do not result in free electrons but just redistribute the photon energies.

$N_{\rm ion}(E_\gamma)$ is the number of final ionizations that the primary-ionization electron generated by a single photon of energy $E_\gamma$ produces. It is simply given by 
\beq
N_{\rm ion}(E_\gamma) = \eta_{\rm ion} (x_{\rm ion}(z))\ E_\gamma \left[ \frac{n_{\rm H}}{n_{\rm A}}\frac{1}{e_{\rm i,H}} + \frac{n_{\rm He}}{n_{\rm A}}\frac{1}{e_{\rm i,He}} \right] = \eta_{\rm ion} (x_{\rm ion}(z))\ \frac{E_\gamma}{\rm GeV}\, \mu
\eeq
in terms of the ionization potential energies of hydrogen $e_{\rm i,H} = 13.7\ {\rm eV}$ and helium $e_{\rm i,He} = 24.6\ {\rm eV}$ and their respective number abundances in the IGM. Here $\mu = 2.35\cdot 10^7$ GeV$^{-1}$ corresponds to the number of ionizations that an electron of 1 GeV would end up causing if it were to release all of its energy in reionizations. 
The factor $\eta_{\rm ion}$ takes into account the fact that only a portion of that energy actually goes into ionizations, the rest causing only heating and atomic excitations. Such fraction depends in turn on $x_{\rm ion}(z)$ itself, as determined by the detailed studies in~\cite{fractionheating, KamionkowskiChen}:
\beq
\eta_{\rm ion}\big(x_{\rm ion} (z) \big) = \frac{1-x_{\rm ion}(z)}{3}.
\label{etaion}
\eeq

The spectral number density of DM-produced photons $\frac{dn}{dE_\gamma}(z)$ present at redshift $z$ is obtained by integrating the fluxes of photons produced at all previous redshifts ($z^\prime$) taking into account, with an absorption factor, the fact that some of them have already deposited their energies at previous redshifts. 
In formul\ae\  
\beq
\frac{dn}{dE_\gamma}(z) = \int_\infty^z dz^\prime \, \frac{dt}{dz^\prime} \, \frac{dN}{dE^\prime_\gamma}(z^\prime) \ \frac{(1+z)^3}{(1+z^{\prime})^3} \cdot A(z^\prime) \cdot \exp \left[ \Upsilon(z,z^\prime,E_\gamma^\prime) \right].
\label{dndE}
\eeq
Here $\frac{dN}{dE^\prime_\gamma}(z^\prime)$ is the spectrum of photons produced at $z^\prime$ by one single annihilation. The factors of $(1+z)^3/(1+z^\prime)^3$ rescale the number densities taking into account the expansion of the Universe.
The absorption coefficient $\Upsilon$ reads 
\beq
\Upsilon(z,z^\prime,E_\gamma^\prime) \simeq - \int^z_{z^\prime} dz^{\prime\prime} \frac{dt}{dz^{\prime\prime}} n_{\rm A} (1+z^{\prime\prime})^3 \sigma_{\rm tot}(E_\gamma^{\prime\prime})  
\label{absorption}
\eeq
where here $E_\gamma^{\prime\prime} = E_\gamma^\prime (1+z^{\prime\prime})/(1+z^{\prime})$. $A(z^\prime)$ represents the rate of DM annihilations per volume. It encodes therefore the information about the density of annihilating DM particles and in particular the halo formation history, that we discuss in the next subsection.

\medskip

As we already anticipated, beside producing ionization, DM annihilations have also the effect of heating the gas. The other important quantity that we need to compute, therefore, is $T_{\rm igm}(z)$ (that also enters in the recombination rates discussed above). It obeys the differential equation~\cite{NS2}
\beq
\begin{split}
\frac{dT_{\rm igm}(z)}{dz} = & \frac{2\, T_{\rm igm}(z)}{1+z} \\
- & \frac{1}{H_0\, \sqrt{\Omega_{\rm M}}\, (1+z)^{5/2}} \left( \frac{x_{\rm ion}(z)}{1+x_{\rm ion}(z)+ 0.073} \frac{T_{\rm CMB}(z) - T_{\rm igm}(z)}{t_{\rm c}(z)}  + \frac{2\, \eta_{\rm heat}(x_{\rm ion}(z))\, {\mathcal E}(z)}{3\, n_{\rm A} (1+z)^3} \right).
\end{split}
\label{dTdz}
\eeq
The first term just corresponds to the usual adiabatic cooling of the gas during the expansion of the Universe. It would lead to $T_{\rm igm}(z) \propto (1+z)^2$. 

The second term accounts for the coupling between the IG gas and the CMB photons, that have a (redshift-dependent) temperature $T_{\rm CMB}$. When the gas is hotter than the surrounding CMB, some of its energy is transferred to the photons and therefore the gas `Compton-cools' down. On the contrary, if the gas is colder than the CMB, it is warmed up. The expression for the term in eq.~(\ref{dTdz}) is obtained by writing the rate of change between the free electrons of the gas and the CMB photons as~\cite{Comptoncooling} $dE_{e \leftrightarrow \gamma}/dt = 4 \sigma_{\rm T} \,{\mathcal U}\, k_{\rm B}\, n_e (1+z)^3 \, (T_{\rm CMB}-T_{\rm igm})/m_e$ and then translating in terms of the rate of change of $T_{\rm igm}$ of all particles in the gas $dE_{e \leftrightarrow \gamma} \to 3/2\ k_{\rm B} n_{\rm tot} (1+z)^3\, dT_{\rm igm}$ (finally using eq.~(\ref{dtdz}) to pass to redshift)~\cite{Comptoncooling2}. In these relations ${\mathcal U} = \varsigma\, T_{\rm CMB}^4$ is the energy density in the CMB blackbody bath (with $\varsigma$ the Stefan-Boltzmann constant~\cite{StefanBoltzmann}) and $m_e$ is the electron mass. Thus in eq.~(\ref{dTdz}) $t_{\rm c} (z) = 3 m_e / (8\, \sigma_{\rm T}\, \varsigma \, T^4_{\rm CMB}(z))$. The various factors of $(1+z)^3$ rescale the number densities with redshift. 
$n_e = x_{\rm ion}(z)\, n_{\rm A}$ is the fraction of free electrons while $n_{\rm tot} = n_e + n_{\rm H^+} + n_{\rm H} + n_{\rm He} = n_{\rm A} (x_{\rm ion}(z) +1 + 0.073)$ contains the number density of all types of relevant particles in the gas, because it is assumed that collisions keep them at the same temperature (helium is here assumed to remain neutral, for simplicity). 

The third term accounts for the heating induced by DM annihilations. As DM injects energy at a rate ${\mathcal E} (z)$, 
the temperature changes at a rate given by $3/2\, k_{\rm B}\, n_{\rm A}(1+z)^3\, dT_{\rm igm}/dt = \eta_{\rm heat} {\mathcal E}$ (then translated into a rate of change with $z$ as usual). Analogously to eq.~(\ref{etaion}), the factor $\eta_{\rm heat}$ expresses the fact that only a portion of the energy goes into heating. We adopt~\cite{fractionheating}
\beq
\eta_{\rm heat} \big(x_{\rm ion}(z)\big) = C \left[ 1- (1-x_{\rm ion}^a)^b \right]
\eeq
with $C = 0.9971$, $a=0.2663$, $b=1.3163$. 
In terms of the quantities introduced above, the total energy deposited per second per volume by the photons in the intergalactic medium at a given redshift $z$ reads 
\beq
{\mathcal E}(z) = \int_0^{m_\chi} dE_\gamma \, \frac{dn}{dE_\gamma}(z) \cdot  n_{\rm A} (1+z)^3 \cdot \sigma_{\rm tot}(E_\gamma)\cdot E_\gamma .
\label{energy deposited}
\eeq

\medskip

Solving numerically the coupled differential equations (\ref{dxdz}) and (\ref{dTdz}) allows to obtain two expressions for $x_{\rm ion}(z)$ (from which the value for $\delta \tau$ in eq.(\ref{tau})) and $T_{\rm igm}(z)$, to be compared with the observational constraints discussed in the Introduction (eqs. (\ref{tauWMAP}) and (\ref{TIGM})). We integrate the equations from $z=600$ to $z=6$.

\subsection{Structure Formation theory}
\label{structformalism}
The annihilation rate per volume at any given redshift can be thought of as the sum of two
parts $A(z) = A^{\rm sm}(z) + A^{\rm struct}(z)$. The former comes from a uniform density field of Dark Matter, to which we refer as ``smooth'', dominant before structure formation at redshifts $z\gtrsim$100, and can be written as
\beq
A^{\rm sm}(z)=\frac{\langle \sigma v\rangle}{2\,m_\chi^2}\, \rho_{\rm DM,0}^2\, (1+z)^6,
\label{smoothannrate}
\eeq
with $m_\chi$ being the mass of the DM particle, $\langle\sigma v\rangle$ the self-annihilation rate,
and $\rho_{\rm DM,0}$ is the ``smooth'' DM density today $\rho_{\rm DM,0}=\Omega_{\rm DM}\rho_c$, $\rho_c$ being the critical density of the Universe today.
As DM collapses into gravitationally bound structures, the rise of local density will provide an increase in
the rate of annihilations averaged over large volumes; such additional contribution from structure formation can be cast in terms of the number of halos of a given mass $M$ to form at a given redshift $z$, and on the DM density distribution inside them, namely
\beq
A^{\rm struct}(z)=\frac{\langle \sigma v\rangle}{2\,m_\chi^2}\int dM \frac{dn}{dM}(z,M)\, (1+z)^3 \int dr\, 4\pi r^2\, \rho_i^2(r,M(z)) .
\label{structannrate}
\eeq

For the halo mass distribution $dn/dM$ we adopt the Press-Schechter formalism \cite{PressSchechter}
\beq
\frac{dn}{dM}(M,z) = \sqrt{\frac{\pi}{2}} \frac{\rho_{\rm M}}{M}\, \delta_{\rm c}\, (1+z) \frac{d\sigma(R)}{dM} \frac{1}{\sigma^2(R)} \exp \left( -\frac{\delta_{\rm c}^2\, (1+z)^2}{2 \sigma^2(R)} \right)
\label{PS}
\eeq
where $\sigma(R)$ is the variance of the density field inside a radius $R$ and $\delta_{\rm c} = 1.28$.\\
We will consider different cases for the most common halo DM profiles $\rho_i(r)$, commenting more about them in Section \ref{discussion}.
The integral on the halo density squared in eq.~(\ref{structannrate}) can be recast in terms of the virial mass of the halo
\beq
M(z)=\frac 4 3 \pi r_{\rm s}^3\, \Delta_{\rm vir}(z)\, \Omega_{\rm M}\, \rho(z)\, c_{\rm vir}^3(M,z) .
\label{virmass}
\eeq
and the DM halo mass $M_{\rm DM}(z)$ obtained by integrating the DM profile up to the cutoff $c_{\rm vir}(M,z)=r_{\rm vir}(M,z)/r_{\rm s}$ (the concentration parameter)
\beq
M_{\rm DM}(z)=\left(\frac{\Omega_{\rm DM}}{\Omega_{\rm M}}\right)M(z)=4\pi r_{\rm s}^3 \rho_{\rm s}(M(z)) \int_0^{c_{\rm vir}(M,z)}x^2 \,f_i(x)\,dx . 
\label{eqDMmass}
\eeq
Here $r_{\rm vir}$ is the virial radius.  The integration variable is defined as $x\equiv r/r_{\rm s}$, $r_{\rm s}$ is the core radius of the given profile,  $\rho_s(r,M(z))=\rho_i(M(z))/f_i(x)$ and $f_i(x)$ is a functional form for the given type of profile. We discuss our choices for $c_{\rm vir}(M,z)$ and $f_i(x)$, and their impact on the final results in Section \ref{discussion}.

$\Delta_{\rm vir}(z)$ is the virial overdensity of the Universe due to the DM clustering at any given redshift (the radius within which the mean energy density in the halo is $\Delta_{\rm vir}(z)$ times the smooth density at the given redshift $\rho(z)=\rho_{\rm c}\Omega_{\rm M}(1+z)^3$),
depends only on the given cosmology and for a flat $\Lambda$CDM universe can be written as~\cite{ullio}
\beq
\Delta_{\rm vir}(z)=\left(\frac{18\pi^2+82(\Omega_{\rm M}(z)-1)-39(\Omega_{\rm M}(z)-1)^2}{\Omega_{\rm M}(z)}\right) ,
\label{deltavir}
\eeq
being a smooth function of the redshift. It is approximately $18 \pi^2$ for large enough redshifts.

\begin{figure}[t]
\begin{center}
\hspace{-0.5cm}
\includegraphics[width=0.335\textwidth]{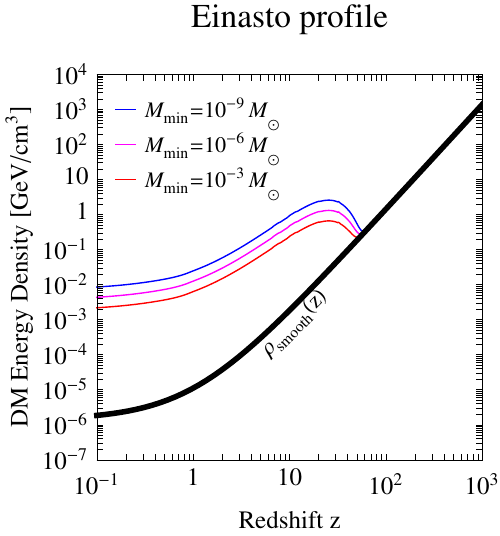}\
\includegraphics[width=0.335\textwidth]{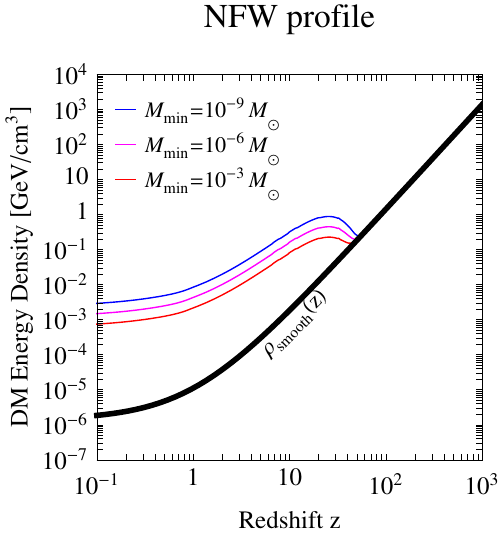}\
\includegraphics[width=0.335\textwidth]{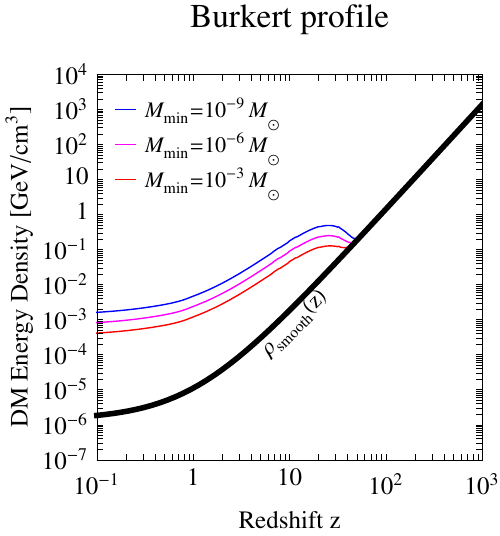}
\caption{\em\label{fig:profileenergydensity} The evolution of the effective DM density $\rho^{\rm eff}_{\rm DM}$ as a function of redshift. Blue, magenta and orange lines refer to $M_{\rm min}$=10$^{-9}$M$_\odot$/10$^{-6}$M$_\odot$/10$^{-3}$M$_\odot$, respectively (from top to bottom). The different panels assume different halo profiles.}
\end{center}
\end{figure}

By defining the concentration function
\beq
F_i(M,z)=c_{\rm vir}(M,z)^3\frac{\int_0^{c_{\rm vir}(M,z)}x^2 \,f_i(x)^2\,dx}{\left(\int_0^{c_{\rm vir}(M,z)}x^2 \,f_i(x)\,dx\right)^2}, 
\label{Fconcentr}
\eeq
we can conveniently recast $A^{\rm struct}(z)$ in terms of a ``boost'' $\mathcal{B}_i(z)$ 
due to the structure formation:
\beq
\mathcal{B}_i(z)=\frac{\Delta_{\rm vir}(z)}{3\,\rho_{\rm c}\Omega_{\rm M}}\int_{M_{\rm min}}^\infty dM\, M\frac{dn}{dM}(z,M)\,F_i(M,z) ,
\label{boostfactor}
\eeq
where $M_{\rm min}$ is the mass of the smallest halos that form, on which we will return below.
In its final form the annihilation rate at any given redshift reads
\beq
A(z)=\frac{\langle \sigma v\rangle}{2\,m_\chi^2}\rho_{\rm DM,0}^2(1+z)^6\left(1+\mathcal{B}_i(z)\right),
\label{annrate}
\eeq
thus allowing us to define an effective, averaged DM density resulting from structure formation, $\rho^{\rm eff}_{\rm DM}(z) = \rho_{\rm DM,0}\, (1+z)^3 \sqrt{1+  \mathcal{B}_i(z) }$  which we plot in figure \ref{fig:profileenergydensity}, for different cases. We discuss it in the following section.

%%%%%%%%%%%%%%%%%%%
%%%%%%%%%%%%%%%%%%%

\section{Discussion}
\label{discussion}

Armed with the formalism above, we are able to compute the total optical depth and the final temperature of the IG gas resulting from DM annihilations. We now discuss its practical implementation.

\subsection{Structure formation parameters}
\label{discussionstructure}

A critical quantity for the integration of eq.~(\ref{boostfactor}) is the concentration parameter $c_{\rm vir}(M,z)$, which can be thought of the (normalized) physical radius of a halo of given mass $M$.
It is usually obtained by the results of numerical simulations, and in particular is found to be inversely proportional to the redshift $z$, namely $c_{\rm vir}(M,z)$=$c_{\rm vir}(M,0)$/$(1+z)$ (Bullock et al. (2001) in \cite{structform}), as the radius of a halo of given mass grows with the redshift as the Universe expands.
We have adopted the $c_{\rm vir}(M,0)$ best fitting a WMAP3 cosmology \cite{Spergel:2006hy},  from \cite{Maccio':2008xb} (Eq. 9).
The core radius $r_{\rm s}(M)$ is instead the radius of the core of a halo of given mass $M$, and its size depends on the chosen profile. In the table at page \pageref{tableprofiles} we give the adopted values of $r_{\rm s}(M)$ for a Milky Way sized halo, and the corresponding energy density $\rho_{\rm s}$.

\medskip

The Dark Matter profiles of the forming halos are assumed to be determined by numerical simulations. Recent, state-of-the-art computations seem to converge towards the so called Einasto profile~\cite{Graham:2005xx, Navarro:2008kc}
\begin{equation}
 \rho_{\rm Ein}(x)=\rho_{s} \exp\left[-\frac{2}{\alpha}\left(x^{\alpha}-1\right)\right],\ \ \alpha=0.17.
   \label{eq:Einasto}
\end{equation}
The Navarro-Frenck-White profile~\cite{Navarro:1995iw} and the isothermal-like Burkert profile~\cite{Burkert}\\[3mm]
\begin{minipage}{0.495\linewidth}
\begin{equation}
 \rho_{\rm NFW}(x)=\rho_{s}\, \frac{1}{x} \frac{1}{\left(1+x \right)^{2}}
   \label{eq:NFW}
\end{equation}
\end{minipage},
\vspace{2mm}
\begin{minipage}{0.495\linewidth}
\vspace{2mm}
\begin{equation} 
   \rho_{\rm Burkert}(x)=\frac{\rho_{s}}{(1+x)(1+ x^{2})}
   \label{eq:isoT}   
\end{equation}
\end{minipage}
represent instead previously standard choices. The shape of the DM profile enters in the concentration function of eq.~(\ref{Fconcentr}). We select the NFW profile as a benchmark, in which case, explicitly, 
\beq
F_{\rm NFW}(c_{\rm vir}) = \frac{c_{\rm vir}^3}{3} \left(1 + \frac{1}{(1+c_{\rm vir})^3} \right) \left(\log(1+c_{\rm vir}) - \frac{c_{\rm vir}}{1+c_{\rm vir}} \right)^{-2}.
\eeq
Analogous, more cumbersome, expressions can be easily calculated for the Einasto and Burkert profile. One actually finds that  the concentration functions for these latter profiles can be parameterized as $F_{\rm Ein} = k_{\rm Ein}^3 F_{\rm NFW}$ with $k_{\rm Ein} \simeq 2.0$ and $F_{\rm Bur} = k_{\rm Bur}^3 F_{\rm NFW}$ with $k_{\rm Bur} \simeq 0.7$, along a large range of $c_{\rm vir}$. We will present results for the three profiles. As one can already infer from the numerical values of the $k_{\rm i}$ coefficients, the adoption of an Einasto(/Burkert) profile will lead to an annihilation rate due to structures (and therefore to constraints on DM) that are roughly a factor of 8 stronger (/3 weaker) with respect to NFW.

\medskip

Finally, another quantity whose determination is somehow non-univocal in literature, and that somewhat affects the final results, is the minimal halo mass to form, $M_{\rm min}$, which we need to set the lower limit for the integration of our eq.~(\ref{boostfactor}). This is usually picked to be the free-streaming mass, $M_{\rm fs}$, associated to a given DM particle mass, and it can be calculated from first principles once the type of interactions of the DM particle is assigned. An expression for $M_{\rm fs}$ is e.g.  in~\cite{Bringmann:2009vf} (Eq. 13), and the kinetic decoupling temperature $T_{\rm kd}$ can be taken from \cite{Chen:2001jz} (Eq. 19), since in the cases in which we are interested particles are typically weakly interacting with baryons. According to the standard paradigm, we obtain $M_{\rm fs} \sim 10^{-6}(/10^{-8}/10^{-11}) M_\odot$ for a WIMP mass $M=100\, {\rm GeV (/1TeV/10TeV)}$.
It has been recently questioned, however, that the minimal halo mass is not $M_{\rm min}=M_{\rm fs}$, but that the first halos collapse in fact at smaller redshift than previously expected, and the typical mass is $M_{\rm min} =10^5 - 10^7 M_{\rm fs}$  \cite{Angulo:2009hf}.

For our final results, we have chosen to present bounds obtained by picking $M_{\rm min} = 10^{-6} M_{\odot}$; we have however performed full calculations by using different values of $M_{\rm min}$, and we find that our constraints become stronger(/weaker) of a factor $\sim$4 by using $M_{\rm min}=10^{-9}M_{\odot}(/10^{-3}M_{\odot}$).

\smallskip

Before concluding this section, let us anticipate now that the bound from the optical depth will actually be insensitive to the parameters of structure formation, as it originates from the epoch when DM had not started collapsing in structures yet. The bound from the temperature, instead, will depend on the choices discussed here.

\subsection{Particle Dark Matter scenarios}
We consider, in a model independent way, Dark Matter particles that annihilate into different primary channels (with a 100\% branching ratio). We use
$${\rm DM}\ {\rm DM} \to e^+e^-, \mu^+\mu^-, \tau^+\tau^-, W^+W^-, b \bar b, t \bar t .$$
Each channel produces different spectra of prompt photons and of electrons+positrons, that we compute with the use of the PYTHIA MonteCarlo code~\cite{PYTHIA} as described in detail in~\cite{CKRS}.\footnote{We stick to the case of ``direct" annihilation of DM particles into a pair of Standard Model particles, i.e. we do not study models in which the annihilation proceeds into some new light mediator state (see for instance~\cite{Pospelov,Arkani, Papucci, Nomura,BertoneTaosoSweden,Schwetz,StrumiaPapucci}). These ``cascade annihilation processes'' lead generically to softer spectra.} We then calculate the spectra of IC scattering photons, produced by the energetic electrons and positrons from DM, as discussed in~\cite{CirelliPanci}. The ambient light on which ICS occurs consists here of the CMB radiation only, and it bears a dependance on the redshift. The sum of the prompt photons and the ICS photons thus computed constitutes the $\frac{dN}{dE^\prime_\gamma}(z^\prime)$ to be plugged in eq.~(\ref{dndE}). 

We scan over a large range of DM masses $m_\chi$, from light Dark Matter (10 GeV) up to very heavy Dark Matter (10 TeV).

For the Dark Matter profiles of the Milky Way today $\rho^{\rm MW}\left( r/r^{\rm MW}_{\rm s} \right)$ we consider the three different models discussed above: Einasto, NFW and Burkert. The values for the parameters $r^{\rm MW}_{s}$ and $\rho^{\rm MW}_{s}$ are given by 
$$ \begin{tabular}{l|cc}
 \label{tableprofiles}
  DM halo model & $r^{\rm MW}_{s}$ in kpc & $\rho^{\rm MW}_{s}$ in GeV/cm$^{3}$\\
  \hline
  NFW \cite{Navarro:1995iw} & 20 & 0.26\\
  Einasto \cite{Graham:2005xx, Navarro:2008kc} & 20 & 0.06\\
  Burkert \cite{Burkert} & 5 & 3.15
 \end{tabular}$$
Note that all profiles are normalized at $\rho^{\rm MW}_\odot = 0.3\ {\rm GeV}/{\rm cm}^3$ at the location of the Earth~\cite{rhoEarth} (see however also~\cite{rhoEarth2}). 
%Changing this value would just rescale all our results by a constant factor. 

\bigskip

In this broad parameter space lie the regions that allow to explain in terms of DM annihilations the CR anomalies mentioned in the Introduction, as we now better detail.

\medskip

{\bf Dark Matter fits to charged Cosmic Ray data.}
The PAMELA satellite~\cite{PAMELA} has reported a significant excess of the positron fraction $e^+/(e^++e^-)$ above the expected smooth astrophysical background and a steep rise above 10~GeV up to at least 100~GeV~\cite{PAMELApositrons}, compatibly with previous less certain hints from HEAT~\cite{HEAT} and AMS-01~\cite{AMS01}.
At the same time, no signal in the $\bar p$ fluxes has been seen, up to the maximal probed energy of about 100 GeV~\cite{PAMELApbar}.
An excess in the flux of $e^++e^-$ has also been reported by the ATIC-2~\cite{ATIC-2} and PPB-BETS~\cite{PPB-BETS} balloon experiments, that in particular found a peak at about 500-800 GeV. The ATIC-4 balloon flight has also confirmed its presence~\cite{ATIC-4}.
Later, this sharp feature has been questioned by the results of the FERMI satellite~\cite{FERMIleptons}: while an excess with respect to the expected background is confirmed, the $e^++e^-$ spectrum has been found to be instead reproduced by a simple power law with index  $-3.04$. The HESS \v Cerenkov telescope, too, has published data~\cite{HESSleptons, HESSleptons2} in the range of energy from 600 GeV up to a few TeV, showing a power law spectrum in agreement with the one from FERMI and eventually a steepening at energies of a few TeV.

We perform the fits to these data\footnote{We consider the PAMELA data for positrons at energies larger than 10 GeV only, where the uncertainty due to solar modulation is not present. We include the systematic error band on the FERMI and HESS datapoints.} with the use of DM generated $e^+,e^++e^-$ and $\bar p$ spectra, as discussed in detail in~\cite{CKRS}: we find the best fit values by scanning over the propagation parameters of charged cosmic rays and over the uncertainties on the slope and normalization of the astrophysical electron, positron and antiproton background. We do not include any galactic boost factor due to substructures within the Milky Way halo. 

The resulting allowed regions on the plane `DM mass' -- `Annihilation cross section' are shown in fig.s~\ref{fig:exclusion1} and~\ref{fig:exclusion2}. 
In order to fit the PAMELA data alone (for positrons and antiprotons), the DM particle has to fall in one of two classes: (a) a DM that annihilates only into leptons (${\rm DM}\, {\rm DM} \to e^+e^-,\mu^+\mu^-,\tau^+\tau^-$), or (b) a DM with a mass around or above a few TeV, that can then annihilate into any channel (i.e. ${\rm DM}\, {\rm DM} \to W^+W^-, ZZ, b\bar b, t\bar t$, light quark pairs and the leptonic channels above) possibly producing anti-proton fluxes at energies above those currently probed by PAMELA. These regions are individuated by green and yellow bands, for 95\% C.L. and 99.999\% C.L., corresponding to $\Delta \chi^2(m_\chi,\langle \sigma v\rangle) \equiv \chi^2(m_\chi,\langle \sigma v\rangle) -\chi^2_{\rm min} \simeq 6$ and $\Delta \chi^2 \simeq 23$, with 2 d.o.f..

The addition of the ATIC (and PPB-BETS) data had pinned down the mass of the DM particles univocally, at about 1 TeV~\cite{CKRS}, and therefore imposed class (a) as the only possibility (see~\cite{CKRS}). This is still the case with the FERMI data replacing the balloon results: the absence of features in the FERMI spectrum requires the DM mass to be somewhat above 1 TeV, but the HESS indication of a cut-off limits it below a few TeV. The regions that fit PAMELA+FERMI+HESS combined are represented in the figures by red and orange areas (for the same confidence levels as above). Notice that the smoothness of the FERMI spectrum forbids a reasonable fit with the ${\rm DM}\, {\rm DM} \to e^+e^-$ channel, that would produce too peaked a feature: the reduced $(\chi^2_{\rm min})_{\rm red}$ turns out to be well above 2 for all DM profiles, so we plot no allowed region.

In all cases, a very large annihilation cross section is needed: of the order of $10^{-23}\, {\rm cm}^3/{\rm sec}$ up to $10^{-20} \,{\rm cm}^3/{\rm sec}$ or more, depending on the mass of the candidate and the annihilation channel. Our results reproduce and agree reasonably well with those of~\cite{CKRS,StrumiaPapucci} (for the cases that overlap).

\medskip

The needed large cross sections  can be justified in specific models in terms of some enhancement mechanism which is effective today but not in the Early Universe, such as a Sommerfeld  enhancement (see~\cite{Sommerfeld,MDMastro,CKRS}, and then~\cite{Arkani,Sommerfeld2,bovy}) due to the exchange of force--carriers at relative low--velocities of the annihilating particles (which has to be much more massive than the carrier). Although the details of the enhancement of the cross--section are model dependent, some general features can be identified: in particular, the enhancement shows an inverse proportionality with the relative velocity of the two particles, and it typically saturates to a maximum value when that is $\beta \lesssim 10^{-4}$. At recombination, $z \sim 1000$, the typical value for self-annihilating DM particles (with masses in the range explored by our analysis) is $\beta\sim 10^{-8}$, and the smooth density field only cools further with the expansion of the Universe. When DM collapses into gravitationally bound structures it soon virializes inside the halo, therefore being heated up by the gravity. However, most of the contribution to the annihilation power at $z \gtrsim 6$ comes from haloes with masses $M \lesssim 10^6M_\odot$, for which the virial velocity is approximately $v_{\rm vir} \sim 10$ km/s; this implies that almost all the annihilation contributing to our signal take place in environments with values of $\beta\sim$10$^{-5}$, thus guaranteeing that the Sommerfeld effect is saturated.

\section{Results: bounds on Dark Matter properties}
\label{results}

\begin{figure}[t]
\begin{center}
\includegraphics[width=0.327\textwidth]{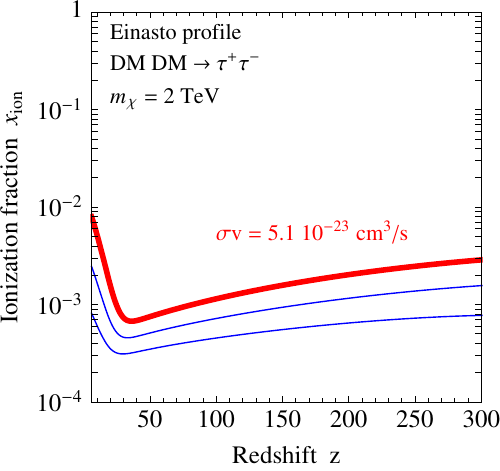}\ \
\includegraphics[width=0.317\textwidth]{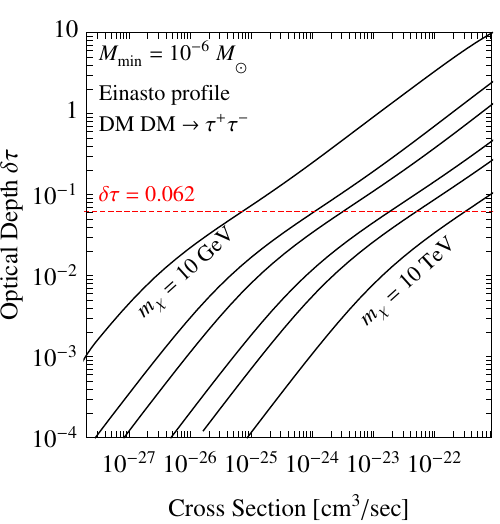}\ \
\includegraphics[width=0.327\textwidth]{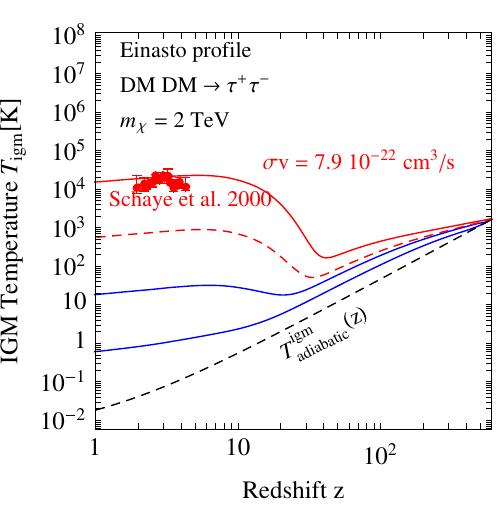}\ 
\caption{\label{fig:results} \em The effects produced by DM annihilations on the observables connected to reionization and heating, for the case of annihilations into $\tau^+\tau^-$. {\em Left panel:} the ionized fraction $x_{\rm ion}$ as function of redshift for $m_\chi = 2$ TeV and for increasing annihilation cross sections $\langle \sigma v \rangle$ = (0.4, 1.4, 5.1) $\cdot$ 10$^{-23}$ cm$^3$/sec (bottom to top). The thick red line individuates the value of the cross section for which the integrated optical depth exceeds the constraints. {\em Central panel:} the integrated optical depth $\delta \tau$ contributed by DM annihilations as a function of the annihilation cross section $\langle \sigma v \rangle$. We plot lines corresponding to $m_\chi$ = (10, 70, 170, 800, 2000, 10000) GeV, left to right.
For a given mass, values of the cross sections larger than those where the horizontal line of the residual optical depth $\delta \tau = 0.062$ is crossed are excluded.
{\em Right panel:} The temperature $T_{\rm igm}$ of the IGM as a function of redshift $z$, for different annihilation cross sections (increasing from bottom to top, the lower blue lines correspond to (0.2, 3.1) $\cdot$ 10$^{-24}$ cm$^3$/sec). The lowermost dotted line shows the adiabatic cooling in absence of DM annihilations. The data points reproduce the measurements of Schaye et al (2000)~\cite{TigmObs2}.  In this example, any annihilation cross section larger  than $7.9 \cdot 10^{-22} {\rm cm^3}/{\rm sec}$ (the uppermost red solid line) would lead to excessive heating of the IGM and is therefore excluded. The red dashed line corresponds to the cross section that already exceeds the $\delta \tau$ bound, which is therefore much more constraining for this example.
}
\end{center}
\end{figure}

Solving numerically the coupled differential equations (\ref{dxdz}) and (\ref{dTdz}) allows to obtain two expressions for $x_{\rm ion}(z)$ (from which the value for $\delta \tau$ in eq.(\ref{tau})) and $T_{\rm igm}(z)$, to be compared with the observational constraints discussed in the Introduction (eqs. (\ref{tauWMAP}) and (\ref{TIGM})).

\medskip

In fig.\ref{fig:results} we show an example of how the DM-generated quantities affect these observables.
Fig.\ref{fig:results}a shows the evolution of the ionized fraction $x_{\rm ion}$ as a function of redshift for increasing annihilation cross sections (bottom up). This refers to a ${\rm DM}\ {\rm DM} \to \tau^+ \tau^-$ annihilation channel and we have chosen $m_\chi = 2$ TeV. One sees that, as soon as the cross section is sizable, the ionized fraction is already quite high at high redshifts, so it will yield a large $\delta \tau$ when integrated according to eq.~(\ref{tau}). 
The red uppermost line individuates the cross section that exceeds $\delta \tau = 0.062$, the value that fills the gap between $\tau(z \le 6) \simeq 0.038$ and the upper bar of the measured $\tau$ in eq.~(\ref{tauWMAP}) (see the discussion in the Introduction).
The fact that, very early, the integral saturates 0.062 just means that the dominant contribution is the one from the annihilation of the smooth DM density field, before structures even start  forming. When they do, at redshift $z \approx 50$, the effect is visible in the plot in a slight increase of $x_{\rm ion}$, that however does not contribute much to the integrated optical depth. For this reason, we expect the bound from $\delta \tau$ to be insensitive to the parameters that govern the history of structure formation.

Parallely, Fig.\ref{fig:results}b shows how the optical depth $\delta \tau$ varies as a function of the annihilation cross section (on the horizontal axis) for different choices of the DM mass $m_\chi$, so that here it is easy to see how the bound arises: for a given mass, the values of the annihilation cross section for which the DM-produced $\delta \tau$ exceeds 0.062 are excluded. For small masses, relatively small DM annihilation cross sections will already produce too much reionization. As the mass increases, the density of DM particles, and therefore the rate of annihilations and the energy injection, decreases. Therefore larger cross sections are admitted without exceeding $\delta \tau = 0.062$.

Notice that, in order to determine the bounds, we impose that DM is the only source of reionization earlier than $z = 6$. As discussed in the Introduction, it is instead astrophysically motivated to believe that quasars and early stars have contributed (it would actually be quite unphysical to believe that astrophysical sources did not contribute at all to reionization), but the size of their contribution is unknown. Including the reionizing impact of astrophysical sources would only reduce the room for DM, leading to stronger constraints. 
We also do not include the contribution to $\delta \tau$ from the small amount of `relic' free electrons left over from CMB formation. This contribution is not well determined: if estimated conservatively at an average $x_{\rm ion, relic} \approx {\rm few} \cdot 10^{-4}$ it contributes (up to $z \simeq 700$) just a small amount $\delta \tau_{\rm relic} \approx 0.01$; but, if larger, it could even provide a sizable portion of the required $\delta \tau$~\cite{SchullVent}, reducing considerably the room for DM and leading again to stronger constraints. Ours are therefore conservative choices.

In fig.\ref{fig:results}c we show how the temperature $T_{\rm igm}$ of the intergalactic medium is affected by DM annihilations, still for the choice of primary annihilation channel ${\rm DM}\ {\rm DM} \to \tau^+ \tau^-$, for the given DM mass $m_\chi = 2\ {\rm TeV}$ and for an Einasto profile. We show $T_{\rm igm}(z)$ for increasing values of the annihilation cross section $\langle \sigma v \rangle$. 
At high redshifts the temperature already departs from the standard adiabatic cooling behavior, as the DM annihilations of the smooth component inject heat in the gas. At a given moment in redshift (corresponding roughly to the formation of the first halos), the temperature rises more abruptly. The larger the cross section, the earlier the effect on the temperature appears and the higher the low-redshift temperature ends up being. 
We impose to $T_{\rm igm}$ not to exceed the 1$\sigma$ error bar of any of the data points in~\cite{TigmObs2}, reproduced in our fig.~\ref{results}c. This individuates a maximum cross section and imposes therefore the bound. We thus see that this bound is regulated by the behavior of the temperature at low redshifts, when DM structures are forming. It will therefore depend on the choices of parameters of structure formation history, and for instance it would be different for a different choice of DM profile, rescaling with the factors discussed in sec.~\ref{discussionstructure}. 
For the example in the figure, that value of the cross section is actually already excluded by the optical depth one discussed above, corresponding to the dashed red line in the plot: the bound from the temperature is definitely subdominant in this case.

\begin{figure}[t]
\begin{center}
\includegraphics[width=0.327\textwidth]{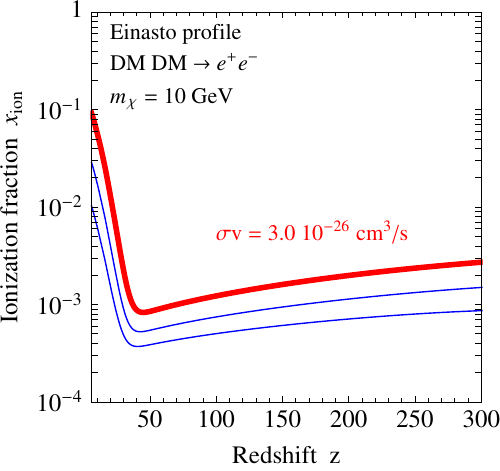}\ \
\includegraphics[width=0.317\textwidth]{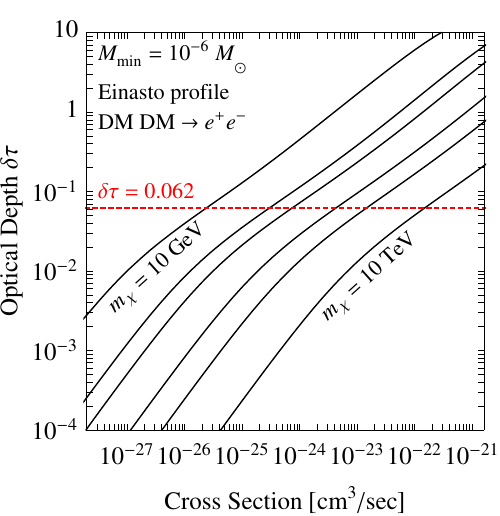}\ \
\includegraphics[width=0.327\textwidth]{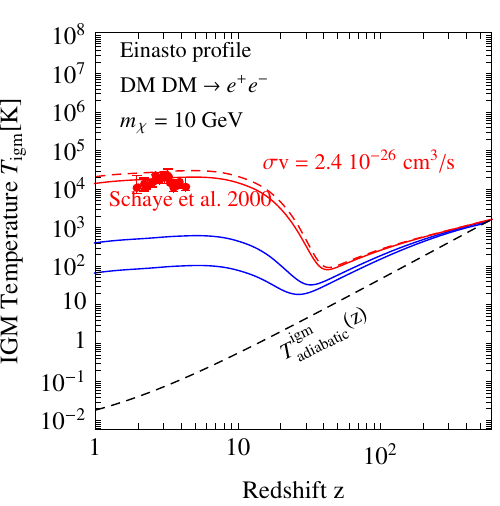}\ 
\caption{\label{fig:results2} \em As in figure~\ref{fig:results}, but for annihilations into $e^+e^-$ and focussing on a small DM mass. {\em Left panel:} the ionized fraction $x_{\rm ion}$ as function of redshift for $m_\chi = 10$ GeV ($\langle \sigma v \rangle$ = (0.2, 0.7, 3.0) $\cdot$ 10$^{-26}$ cm$^3$/sec, bottom to top). {\em Central panel:} the integrated optical depth $\delta \tau$. {\em Right panel:} The temperature $T_{\rm igm}$ of the IGM as a function of redshift $z$. In this example, the maximum annihilation cross section (2.4 $\cdot$ 10$^{-26}$ cm$^3$/sec, the uppermost red solid line) leads to very significant heating of the IGM. The red dashed line corresponds to the cross section that exceeds the $\delta \tau$ bound, which is therefore a bit less constraining for this example.
}
\end{center}
\end{figure}

Given the (gently) decreasing shape of $T_{\rm igm}$ at low redshifts, the highest redshift point in eq.~(\ref{TIGM}) is often the most stringent. Analogously to the discussion above, we stress that this choice is a conservative one: in principle, constraints could be made tighter by choosing a lower allowed maximum value, as it would be the case if one assumes that astrophysical sources are mainly responsible for gas heating and therefore a reduced room is left for DM. 

\medskip

Fig.\ref{fig:results2} shows the same quantities for the case of ${\rm DM}\, {\rm DM} \to e^+e^-$ annihilations and focussing on a small DM mass of 10 GeV. In this case, the ionized fraction $x_{\rm ion}$ behaves similarly to the case in fig.\ref{fig:results}: it shows a more abrupt increase in the epoch of structure formation but the $\delta \tau$ is saturated also in this case by the high redshift part. The temperature, however, reaches higher values than before. Actually, the cross section for which the data points are exceeded is slightly smaller than the one for which $\delta \tau$ is saturated, so in this extreme case the temperature bound is slightly stronger.

\bigskip

\begin{figure}[p]
\begin{center}
\hspace{-8mm}
\includegraphics[width=0.333\textwidth]{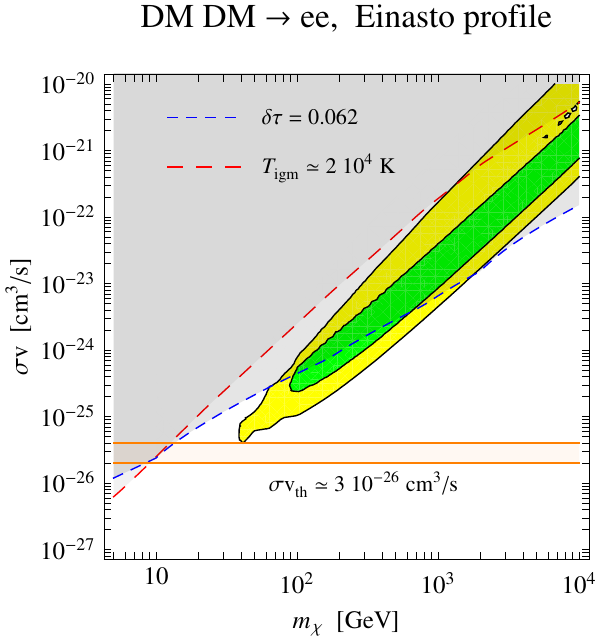}\
\includegraphics[width=0.333\textwidth]{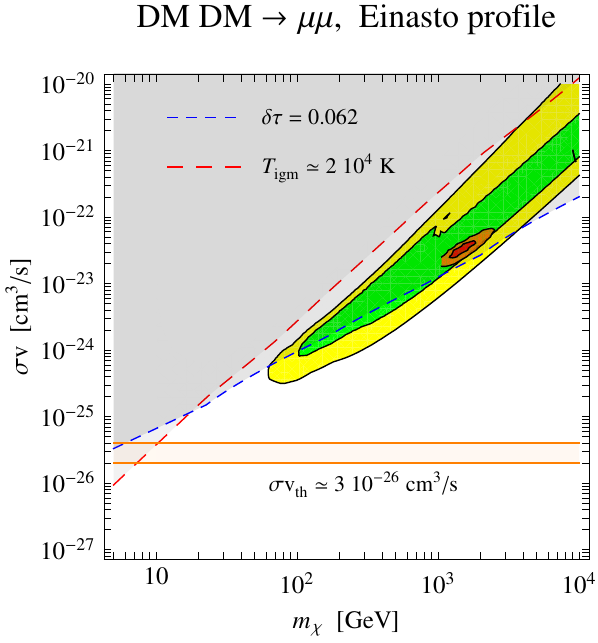}\
\includegraphics[width=0.333\textwidth]{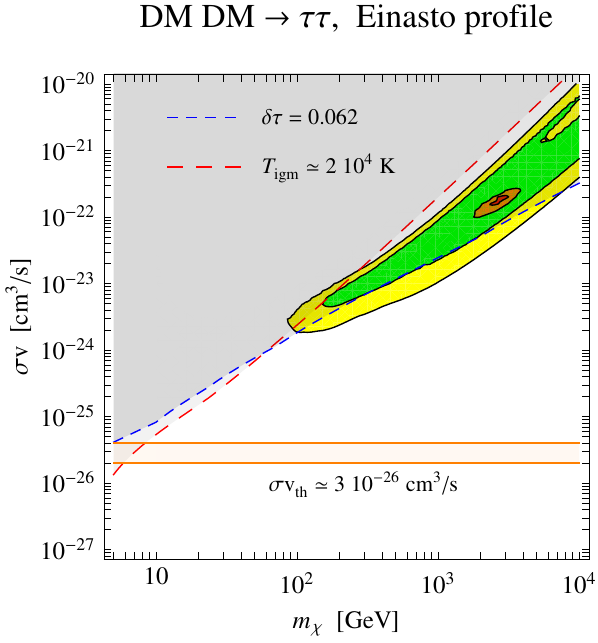}\\[2mm]
\hspace{-8mm}
\includegraphics[width=0.333\textwidth]{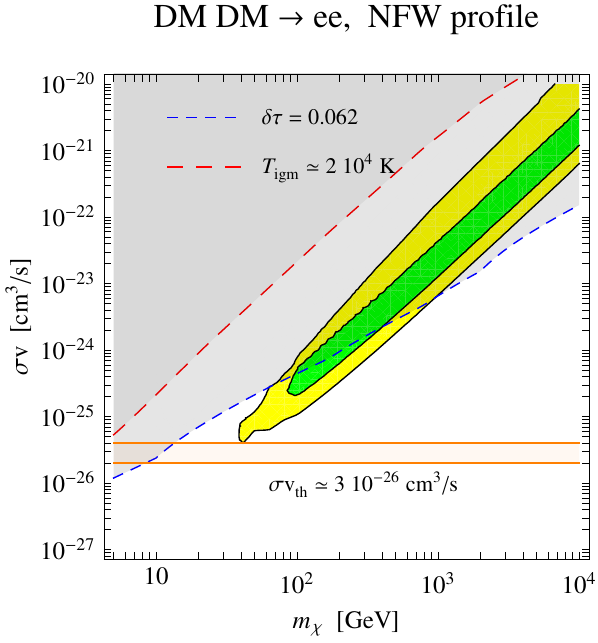}\
\includegraphics[width=0.333\textwidth]{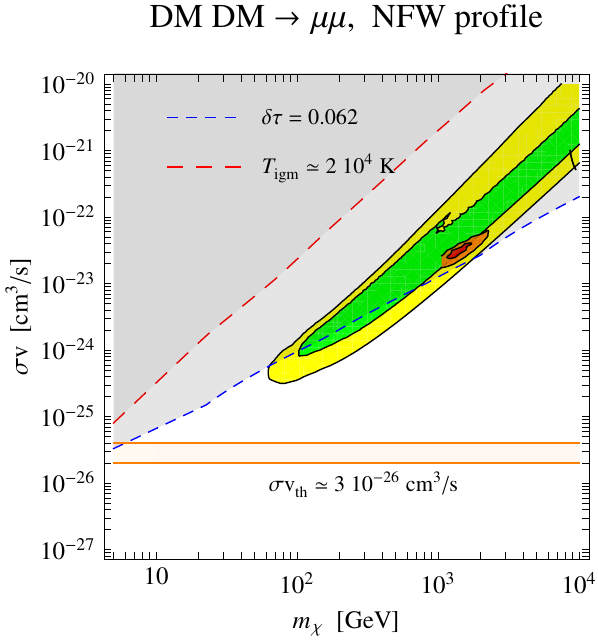}\
\includegraphics[width=0.333\textwidth]{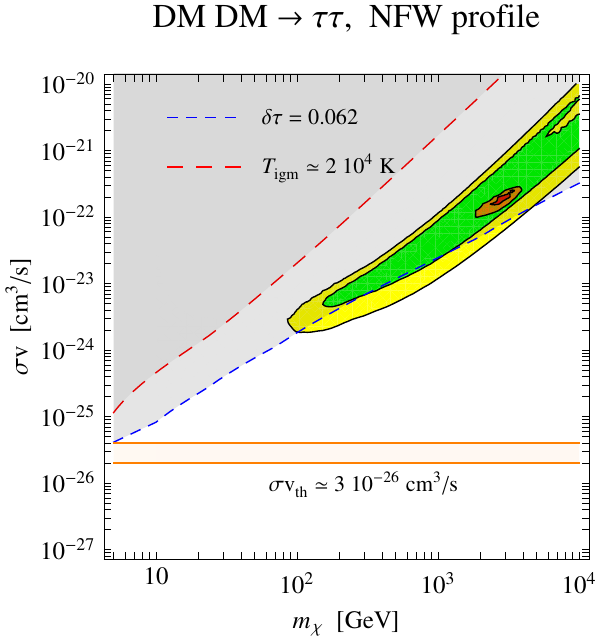}\\[2mm]
\hspace{-8mm}
\includegraphics[width=0.333\textwidth]{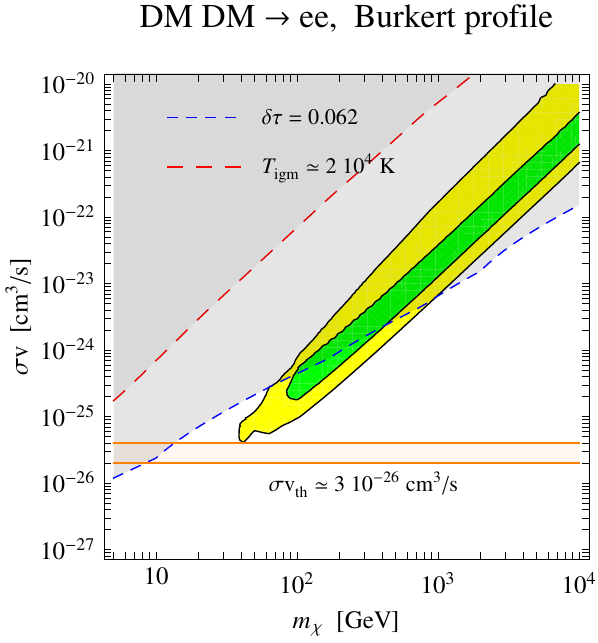}\
\includegraphics[width=0.333\textwidth]{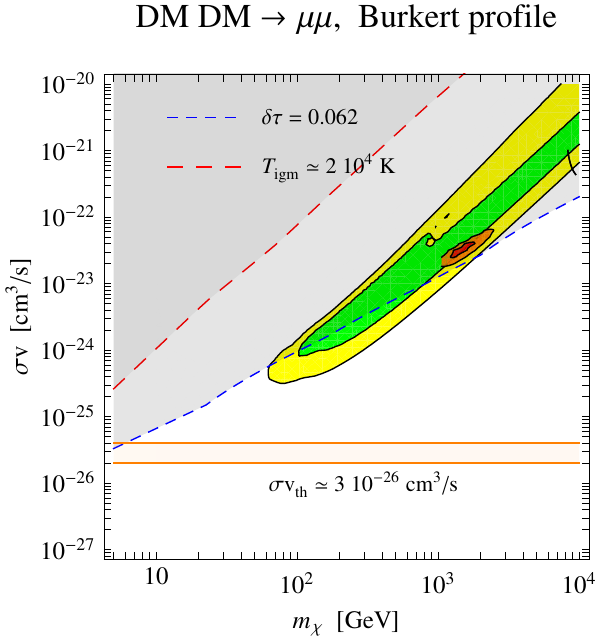}\
\includegraphics[width=0.333\textwidth]{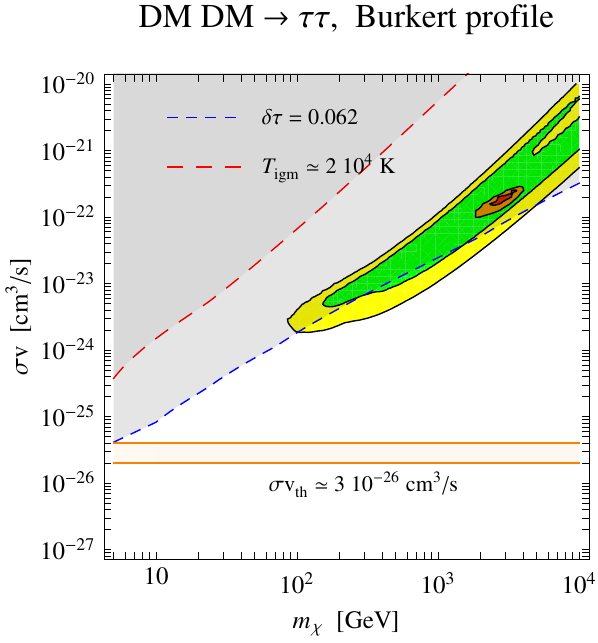}
\caption{\em\label{fig:exclusion1} 
The regions on the parameter space `DM mass' -- `Annihilation cross section' that are excluded by the reionization and heating bounds. The first column of panels refers to DM annihilations into $e^+e^-$, the second into $\mu^+\mu^-$ and the third into $\tau^+\tau^-$; the three rows assume respectively an NFW, an Einasto and a Burkert profile. Each panel shows the exclusion contour due to exceeding the optical depth (blue short dashed line) and the exclusion contour imposed by excessive heating of the intergalactic gas (red long dashed line). We also report the regions that allow to fit the PAMELA positron data (green and yellow bands, 95 \%  and 99.999 \% C.L. regions) and the PAMELA positron + FERMI and HESS data (red and orange blobs, 95 \%  and 99.999 \% C.L. regions). The horizontal orange band indicates the typically preferred value for the thermal annihilation cross section.}
\end{center}
\end{figure}

\begin{figure}[p]
\begin{center}
\hspace{-8mm}
\includegraphics[width=0.333\textwidth]{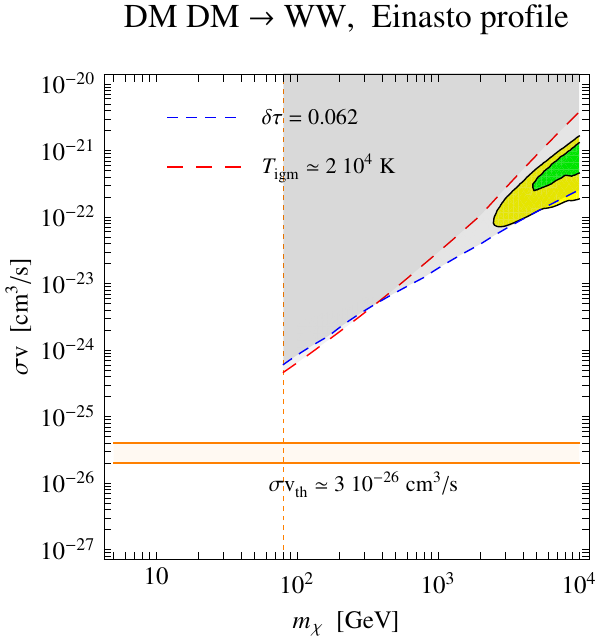}\
\includegraphics[width=0.333\textwidth]{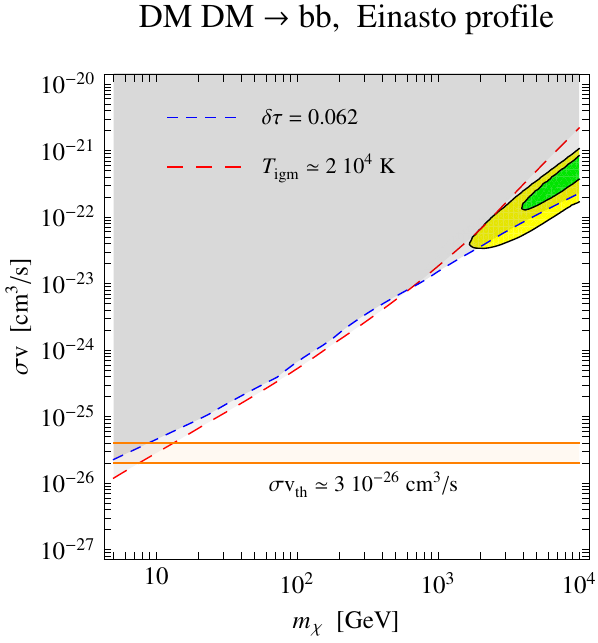}\
\includegraphics[width=0.333\textwidth]{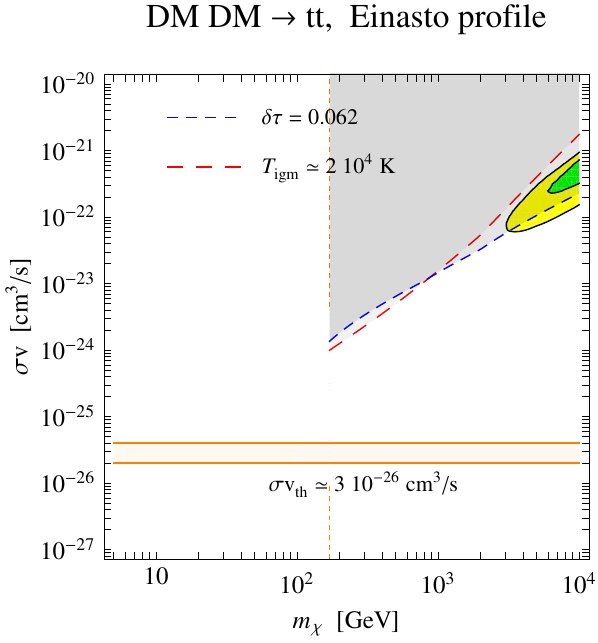}\\[2mm]
\hspace{-8mm}
\includegraphics[width=0.333\textwidth]{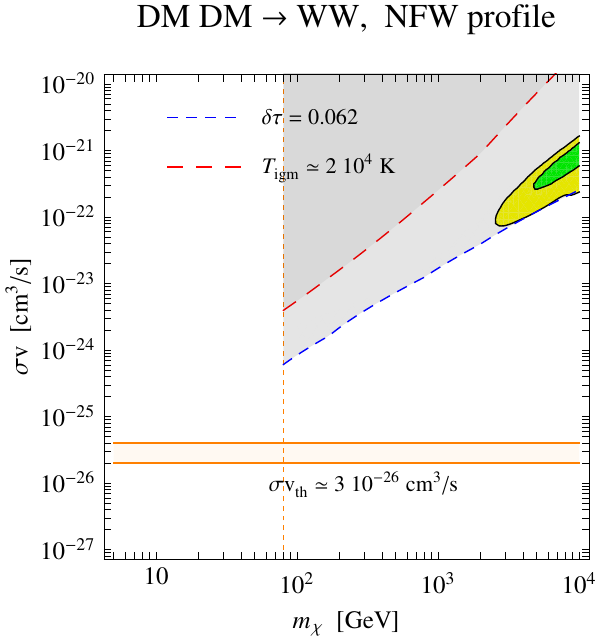}\
\includegraphics[width=0.333\textwidth]{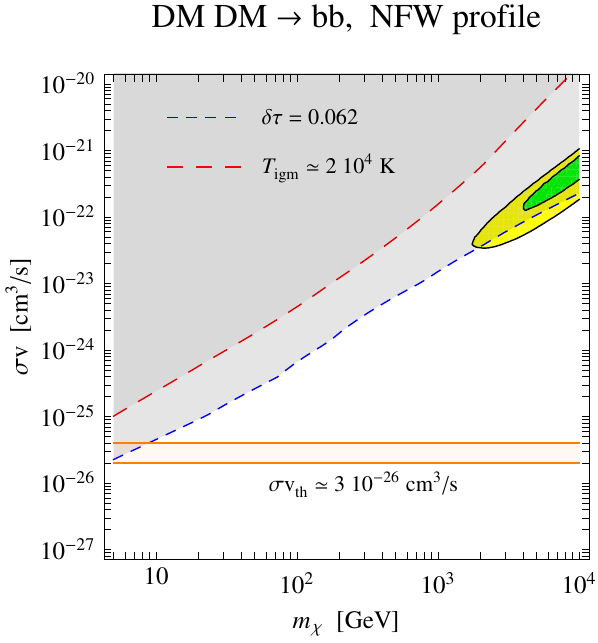}\
\includegraphics[width=0.333\textwidth]{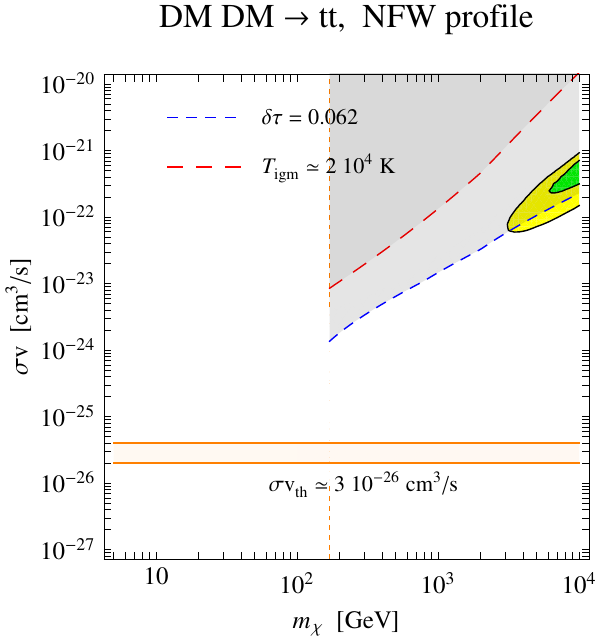}\\[2mm]
\hspace{-8mm}
\includegraphics[width=0.333\textwidth]{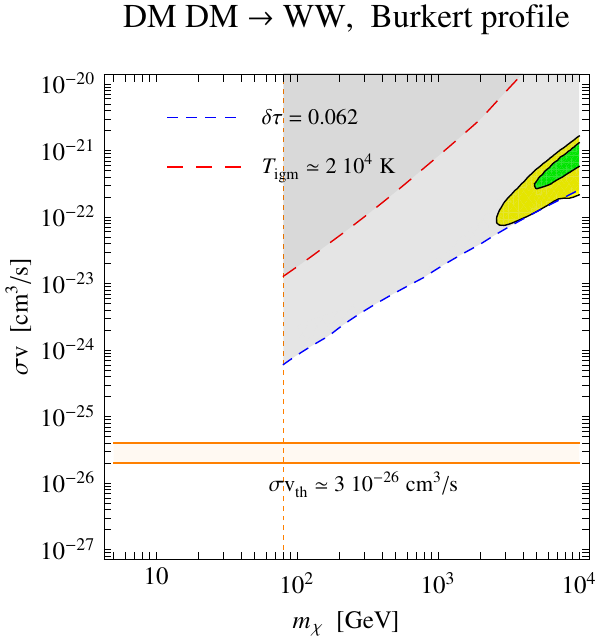}\
\includegraphics[width=0.333\textwidth]{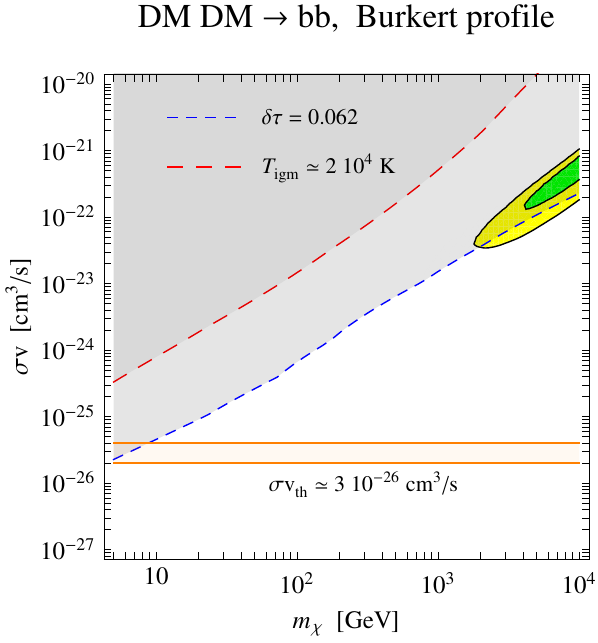}\
\includegraphics[width=0.333\textwidth]{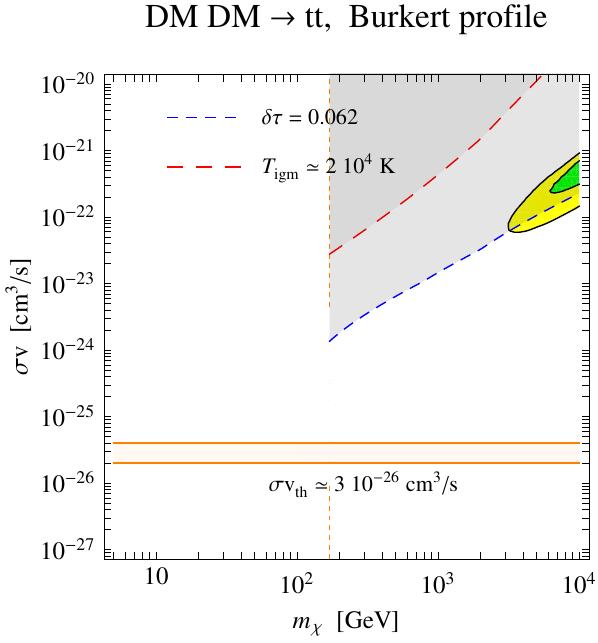}\\[2mm]
\caption{\em\label{fig:exclusion2} 
As in fig.\ref{fig:exclusion1}, but for $W^+W^-$, $b \bar b$ and $t \bar t$ annihilation channels. Since a DM particle fitting the PAMELA data has to be multi-TeV for these channels, the green/yellow bands are confined to large masses. There is no possibility to fit the FERMI and HESS data in these channels. The vertical cut indicates the kinematic threshold for the production of the primary annihilation particles.}
\end{center}
\end{figure}

In fig.s \ref{fig:exclusion1} and \ref{fig:exclusion2} we show the regions on the plane `DM mass $m_\chi$' vs `Annihilation cross section $\langle \sigma v \rangle$' that are excluded on the basis of the optical depth and temperature constraints discussed above, for the different primary annihilation channels that we consider and for different assumed DM halo profiles (used consistently for the forming halos and for our galaxy today).  These plots are produced assuming $M_{\rm min} = 10^{-6} M_\odot$. The main results are that:
\begin{itemize}
\item[$\diamond$] Large portions of the regions individuated by PAMELA and FERMI+HESS are ruled out. For instance, for the case of annihilations into $\tau^+\tau^-$ or $\mu^+\mu^-$, the entire PAMELA and FERMI+HESS region at about 2 to 3 TeV is excluded by the optical depth bound.  
\item[$\diamond$] For small DM masses, the value of the thermal annihilation cross section $\langle \sigma v \rangle_{\rm thermal} \simeq 3 \cdot 10^{-26}\, {\rm cm}^3/{\rm sec}$ starts to be touched by the exclusion contours. The bound from the temperature is most stringent for the $e^+e^-$ primary annihilation channel for the Einasto profile, suggesting  $m_\chi \gtrsim 10$ GeV. For the other profiles or channels both bounds constrains smaller masses.
\end{itemize}
We also comment on the following features. 
\begin{itemize}
\item[-] Changing the DM profile does not affect the bounds from $\delta \tau$, consistently with the fact, discussed above, that it is determined by the ionized fraction before structures even form. The constraint from the temperature instead scales up (/down) by a factor of  $\sim$8 (/3) moving from NFW to Einasto (/Burkert), consistently with the discussion in sec.~\ref{discussion}. Similarly, changing the value for the minimal mass to $10^{-9}M_\odot \ (10^{-3}M_\odot)$ would rescale down (up) the constraints from the temperature by roughly a factor of 4, as discussed in sec.~\ref{discussion}.
\item[-] Changing the primary annihilation channel changes the spectrum of the annihilation products and the fraction of the total energy (corresponding to the rest mass) that is injected in the form of primary annihilation electrons (that make ICS and transfer their energy to photons) and prompt photons. These are the only species that can be absorbed by the IGM and cause ionization, as discussed above. For instance, ${\rm DM}\, {\rm DM} \to \mu^+\mu^-$ or the hadronic channels inject less energy into $e^+e^-\gamma$ because neutrinos and protons carry away a part of the energy. Indeed, the bound rescales up for these channels. The precise ratios depend on the DM mass, the evolution with redshift and the shape of the spectrum (see also~\cite{Slatyer}).
\item[-] The reionization bounds from $\delta \tau$ scales approximately with $\langle \sigma v \rangle/m_\chi$, for all channels and profiles. This behavior is not of immediate understanding, but can be traced back to the energy that DM annihilations inject in the IGM $\propto n_{\rm DM}^2 \langle \sigma v \rangle\,  m_\chi = \rho_{\rm DM}^2 \langle \sigma v \rangle/m_\chi $, to the dependence of the spectrum of $e^+e^-\gamma$ from annihilation on the energy (to be rescaled by redshift) and so ultimately to the fact that the integral in eq.~(\ref{dndE}) bears a global weak dependence on the redshift during the smooth DM period. When instead structures start forming, the dependence on $z$ in the boost factor modifies this scaling behavior. The bound from the temperature (that originates from this later phase, as said) scales more steeply on the plane $m_\chi - \langle \sigma v \rangle$. In the case of the hadronic channels and for not too large masses, however, the different shape of the $e^+e^-\gamma$ spectra partly compensates, and the scaling resembles the one linear  with $\langle \sigma v \rangle/m_\chi$.
\end{itemize}

\section{Conclusions}
\label{conclusions}

We have computed the constraints on Dark Matter annihilations imposed by the reionization and the heating of the intergalactic medium. 

We have calculated the flux of energy injected from DM annihilation, which results in ionization and heating of the  intergalactic medium, {\it including} the important effect of Inverse Compton Scattering photons produced by the energetic $e^\pm$ from DM on the CMB. We have followed the evolution of the smooth DM density and the formation history of DM halos from redshift of a few hundred to today, and solved the evolutions of the population of free electrons (which determines the optical depth of the Universe $\tau$) and of the temperature of the intergalactic gas $T_{\rm igm}$, comparing them with the respective observational measurements. 

\medskip

Fig.s \ref{fig:exclusion1} and \ref{fig:exclusion2} show our main results, in terms of excluded regions on the plane `DM mass' vs `Annihilation cross section'. We have considered several primary annihilation channels and several choices for the DM distribution profiles (consistently used for the forming halos and for our galaxy today). 

We have found that large portions of the regions that allow to fit the PAMELA and FERMI+HESS CR excesses in term of DM annihilations are ruled out by the optical depth bound. For instance, the entire PAMELA and FERMI+HESS region for the $\tau^+ \tau^-$ or $\mu^+ \mu^-$ case around $m_\chi = 2$ TeV is excluded.

We also found that DM particles with small masses, of the order $\mathcal{O}$(10 GeV), tend to produce too much heating even with thermal annihilation cross sections $\simeq 3 \cdot 10^{-26}\, {\rm cm}^3$/sec).\footnote{This is in analogy with the findings of previous works~\cite{previous}, where however DM masses of the order of MeV were under consideration. The inclusion of the ICS photons from DM increases here the energy deposition by several orders of magnitude and so increases significantly the DM mass at which there can be an impact.} The bounds are most stringent for the leptonic primary annihilation channels and for an Einasto profile, suggesting  $m_\chi \gtrsim 10$ GeV in case of an $e^+e^-$ channel. They become weaker for other annihilation channels (as less energy is injected into radiation that can be absorbed by the IGM) and DM profiles.

\medskip
  
In all the parameter space that we studied the dominant contribution to the optical depth is generated by the annihilations of the smooth DM field, before structures even start forming, and so the constraints on DM from this quantity are essentially insensitive to the uncertainties on the structure formation parameters. 
The increase in the temperature originates instead mainly from annihilations during the period of DM clustering, so the constraints are somewhat sensitive to the history of such clustering. The bound from the optical depth is generally stronger but the one from the temperature can become competitive for small masses and for the hadronic channels, in the case of an Einasto profile. 

We stress that these constraints are produced by imposing to the $\delta \tau$ and to the $T_{\rm igm}$ induced by DM not to exceed the 1$\sigma$ error bars of the respective measurements and are derived assuming no competing contribution of optical depth and heat from ordinary astrophysical sources. Including these contributions, with a specific prescription for their impact, would reduce the room for the DM contribution and make the bounds much stronger.

\medskip

The results of our analysis are compatible with Galli et al.~\cite{Galli} who included
the contribution of the only smooth DM density field; their MonteCarlo
analysis on several CMB parameters, clearly dominated by the constrain on
$\tau$ from the WMAP5 data, is however parametrized in terms of an efficient
coupling between the DM shower induced and the IGM gas, $f$. In such
formalism, the parameter $f$ has been computed for a wide range of
secondaries by Slatyer et al~\cite{Slatyer}, that included all the relevant processes we
have self consistently implemented in our analysis. All the results
%, including those by Hurtzi et al. 
are consistent. The actual bounds agree within a numerical factor of a few. Here we have presented a separate estimate of the contribution to $\tau$ given by the smooth and structure boosted annihilations, finding in particular that the smooth one is dominant.

\medskip

For the regions that are still below the bounds but where the effect of DM is already sizable, it would be interesting to investigate whether the resulting ionized and warm intergalactic gas can modify the process of the formation of stellar objects. This kind of analysis is however beyond the scope of our study.

Finally, notice that we have not treated the case of decaying Dark Matter, which has been also recently proposed as an interpretation of the cosmic ray data (see~\cite{decay} for model independent studies and for references). In this case, the relevant quantities are the DM mass and lifetime (as opposed to the annihilation cross section) and the DM energy injection is proportional to the first power of the DM density $\rho_{\rm DM}$ instead of the square. Effects are therefore expected to be generically milder. 
Ref.~\cite{KamionkowskiZhang}, which studies the contribution of the smooth DM density field, finds lower bounds on the lifetime that remain somewhat below the values required by current CR data. 
For a diffuse contribution like the one we are studying, and in the case of decaying DM,  the energy injection averaged over the whole Universe depends on the total mass content, and will therefore be unaffected by clustering due to structure formation. A dedicated study would however be necessary to address this issue more quantitatively in our formalism which, as opposed to the one in Ref.~\cite{KamionkowskiZhang}, treats explicitly the coupling between the DM-induced shower and the IGM gas.

%Finally, notice that we have not treated the case of decaying Dark Matter, which has been also recently proposed as an interpretation of the cosmic ray data (see~\cite{decay} for model independent studies and for references). In this case, the relevant quantities are the DM mass and lifetime (as opposed to the annihilation cross section) and the DM energy injection is proportional to the first power of the DM density $\rho_{\rm DM}$ instead of the square. Effects are therefore expected to be generically milder. Ref.~\cite{KamionkowskiZhang} indeed finds lower bounds on the lifetime that remain somewhat below the values required by current CR data. A dedicated study would however be necessary to address this issue more quantitatively in our formalism.

\paragraph{Note Added.} As this work was in the final stages of completion, on the arXiv appeared ref.~\cite{Raidal}, that also discusses the bound from the optical depth, for leptonic DM annihilation channels and masses above 100 GeV. Our conclusions are consistent and in agreement with theirs, for the cases that overlap.

\paragraph{Acknowledgements}

\noindent We thank Andrea Ferrara, Aravind Natarajan and Lidia Pieri for useful discussions.
The work of P.P. is supported in part by the International Doctorate on AstroParticle Physics (IDAPP) program. 
We thank the EU Marie Curie Research \& Training network ``UniverseNet" (MRTN-CT-2006-035863) for support.
We are grateful to \myurl{www.berenice.tv}{Berenice Sarl} and to Anna Arena for computational support.

\bigskip
\appendix

\footnotesize
\begin{multicols}{2}
  
\end{multicols}

\end{document}